# Monte Carlo tomographic reconstruction in SPECT
# Impact of bootstrapping and number of generated events

Z. El Bitar[1], I. Buvat[2], V. Breton[1], D. Lazaro[2] and D. Hill[3]

[1] Laboratoire de Physique Corpusculaire, CNRS/IN2P3, Université Blaise Pascal, 24 Avenue des Landais, 63177 Aubière Cedex 1, France

[2] INSERM U678, CHU Pitié-Salpêtrière, 91 Boulevard de l'Hôpital, 75634 Paris Cedex 13, France

[3] ISIMA/LIMOS UMR 6158, Computer Science and Modeling Laboratory, Université Blaise Pascal, 24 Avenue des Landais, BP – 10125 ,63173 Aubière Cedex, France

## Abstract:

*In Single Photon Emission Computed Tomography (SPECT), 3D images usually reconstructed by performing a set of bidimensional (2D) analytical or iterative reconstructions can also be reconstructed using an iterative reconstruction algorithm involving a 3D projector. Accurate Monte Carlo (MC) simulations modeling all the physical effects that affect the imaging process can be used to estimate this projector. However, the accuracy of the projector is affected by the stochastic nature of MC simulations. In this paper, we study the accuracy of the reconstructed images with respect to the number of simulated histories used to estimate the MC projector. Furthermore, we study the impact of applying the bootstrapping technique when estimating the projector.*

## Introduction:

Single Photon Emission Computed Tomography (SPECT) is an imaging modality appropriate for visualizing functional information about a patient's specific organ or a body system. A radio-pharmaceutical, which is a pharmaceutical labeled with a radioactive isotope, is administrated to the patient. The radiopharmaceutical is chosen as a function of the organ to be studied. For example, Iodine-131 is appropriate for thyroid imaging. In case of SPECT, the radio-pharmaceutical emits single gamma rays isotropically. Emitted gamma photons are then detected in specific directions using a rotating gamma camera. Rays detected in a given direction yield a projection. At the end of an acquisition process, a set of two-dimensional (2D) projections is available to reconstruct the three-dimensional (3D) radio-pharmaceutical distribution within the body. In the absence of attenuation and scatter of the emitted gamma rays, each point of a projection corresponds to the sum of photons emitted by the source along a straight line (Fig. 1); the detected signal intensity is then:

$$I = \int_L f(x,y)du$$

where I is the detected signal, f(x ,y) the concentration of the radio-pharmaceutical at (x,y) and L represents the line perpendicular to the detector (the gamma camera).

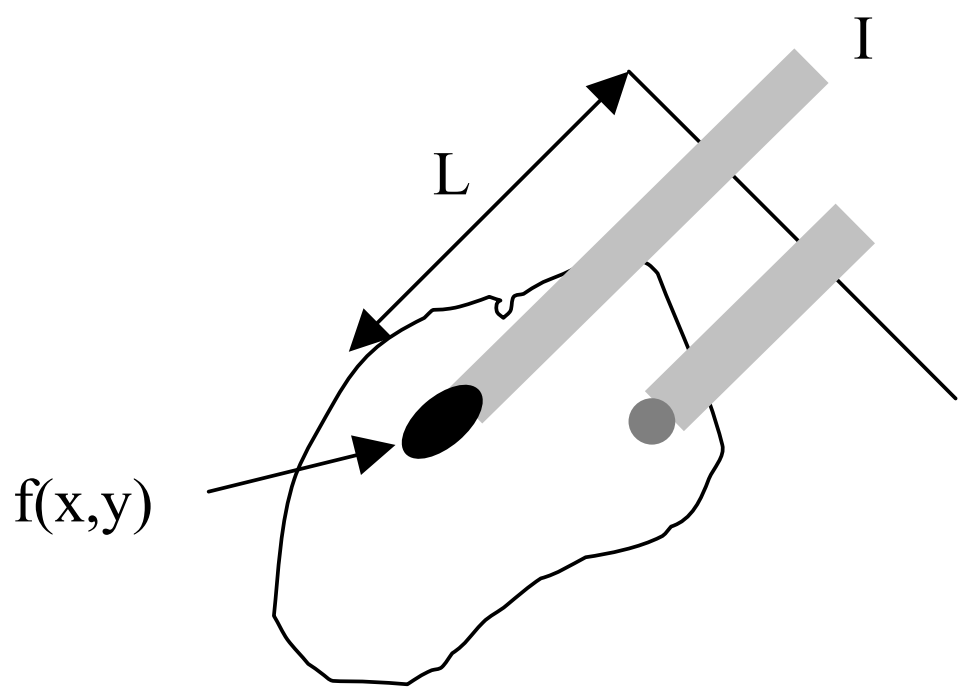

**Figure 1: Detection of a projection in SPECT**

The aim objective is to calculate the radio-pharmaceutical concentration f(x,y) at each point (x,y) knowing a set of projections I along different directions. This is feasible by inverting the Radon transform, thus reconstructing transaxial slices assuming that photons emitted from a transaxial slice are detected within a single line of each 2D planar projection.

A 3D radio-pharmaceutical distribution f(x,y,z) can then be reconstructed as a set of 2D reconstruction of f(x,y) radio-pharmaceutical distribution. The main problem with this approach is that signal measured in projections is affected by physical effects such as scatter, attenuation and detector response. Therefore, photons emitted from a transaxial slice may be detected not only in the projection line facing the slice but also in the neighboring slices. It has been shown that, in realistic configurations, the percentage of such photons is not negligible [Munley et al. 1991]. Furthermore, due to attenuation of the photons within the patient, not all emitted photons reach the detector. Those effects dramatically degrade the reconstructed image. When writing the reconstruction problem in its form where projections and image to be reconstructed are sampled, iterative reconstruction algorithms can be used. This makes it possible to account for most degrading physical effects affecting the projections, by considering a projector modeling these physical effects. Using iterative reconstruction, a 3D radio-pharmaceutical distribution f(x,y,z) can be reconstructed either as a set of 2D f(x,y) distributions, each f(x,y) distribution resulting from a 2D reconstruction involving a 2D projector, ignoring 3D degrading effects, or better, by a fully 3D reconstruction involving a 3D projector accounting for 3D physical effects.

Monte Carlo (MC) simulations have recently been proposed to calculate a 3D projector, in a method called F3DMC (Fully 3 Dimensions Monte Carlo) [Lazaro et al. 2004]. However, the accuracy of Monte Carlo simulations may be adversely affected by the random number generators, seeds used for these generators, number of random drawing, or correlations between simulations. In that context, we tested the impact of the number of drawing and the value of bootstrapping to reduce variance of the projector element estimates [Cheng et al. 2001], hence hopefully improve the quality of the reconstructed images.

## Material and Methods

## 1. Monte Carlo simulations

MC simulations were performed using the software package GATE [Jan et al. 2004]. A cylindrical water phantom including 5 rods (diameters from 4.8 to 11.1 mm) and a bony rod (diameter of 12.7 mm) (Fig 2) was considered. Rod-to-background Tc99m activity concentration ratio between any of the five smallest rods and the background was set to 4. The white rod represents the bony rod.

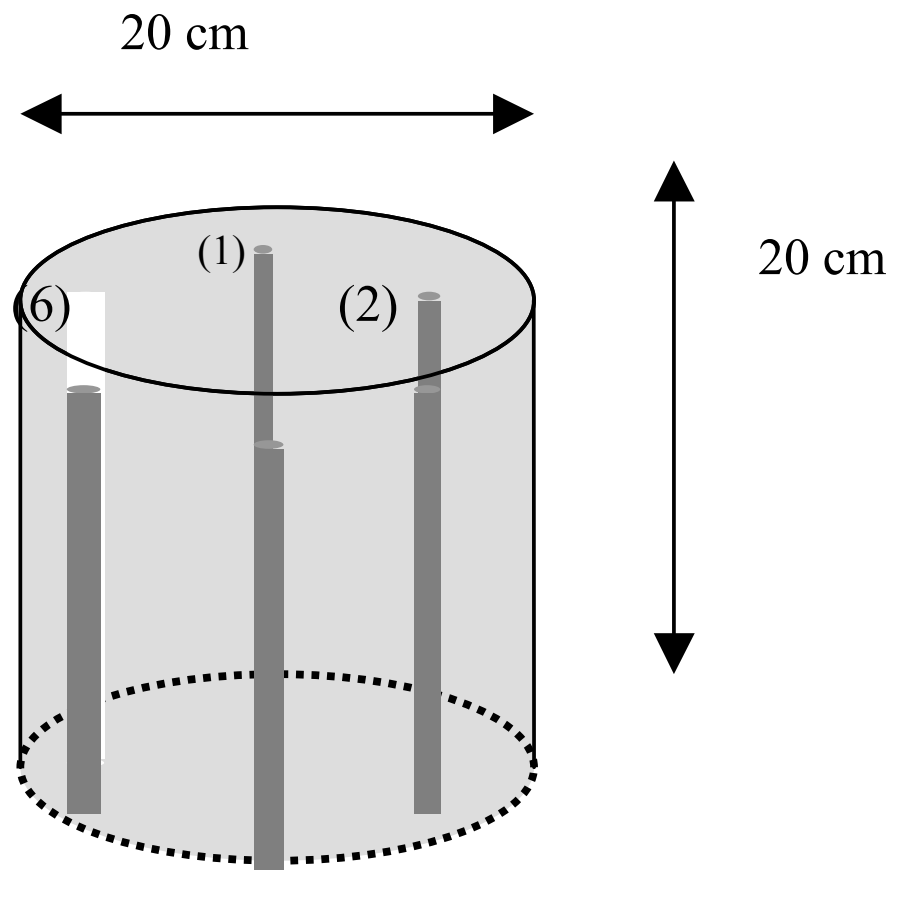

**Figure 2. Simulated phantom**

The simulated camera modeled an AXIS-Philips equipped with a LEHR (Low Energy High Resolution) collimator. 64 projections of 64x64 pixels (pixel dimension= 3.125mm) were simulated and photons detected in the 126-154 keV energy window were considered.

The projector needed for F3DMC was calculated by simulating a uniform activity distribution in the phantom propagation media. The 3D object to be reconstructed was sampled into 64x64x64 voxels (voxel dimension = 3.125mm), yielding a projector with at most 64^6 elements. In practice, only non-zero elements were stored. Given the projections and the projector, the 3D object was reconstructed using F3DMC, by solving the linear system using a Maximum Likelihood Estimation Method (MLEM) algorithm [Miller et al. 1985].

## 2. Image assessment

In order to assess the quality of the reconstructed images (Fig. 3. (b)), rod-to-background activity ratios were measured and compared with reference values. Reference values were obtained by considering the sampled activity distribution that was actually simulated (Fig. 3. (a))

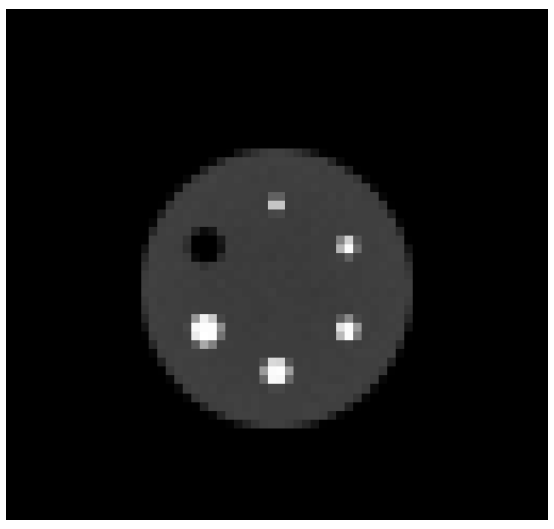

(a)

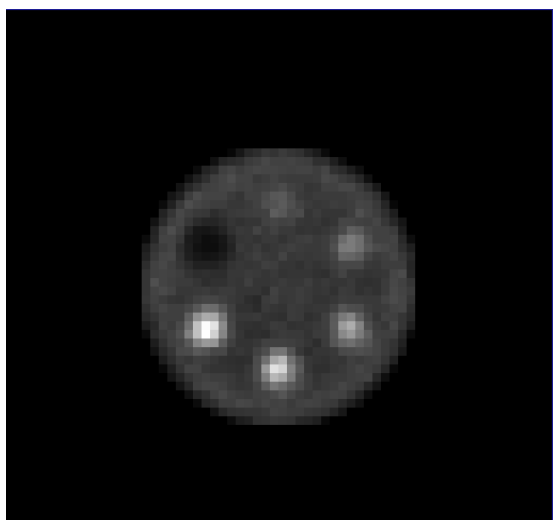

**(b)**

**Figure 3: (a) Reference image; (b) Reconstructed image**

## 3. Statistics used for projector estimate

An accurate projector can be obtained by generating a large number of events in MC simulations, to reduce the standard deviation of each value within the projector. Given that the maximum duration of each simulation was 24 hours, we performed 462 separate sub-simulations, supposed to be independent using the Independent Sequence (IS) technique, i.e. initializing the same generator with ‘n’ different seeds [Codington 1996]. The resulting projector involved in F3DMC method was calculated with 74 billion emitted photons from which 16 million photons were detected in the 126-154 keV window. Each element $P_{ij}$ of the projector was defined as the probability that a photon emitted from a voxel i is detected in the pixel j (1)

(1) $$Pij = \frac{Nij}{Ni},$$

where Nij is the number of photons emitted from voxel i and detected in pixel j, and Ni is the number of photons emitted from voxel i.

## 4. Impact of the number of simulated events on the reconstructed image quality

The principal problem faced while using MC techniques is the long duration of simulations making it difficult to have a large number of statistical histories used in the calculation of the projector. We therefore studied the change of the activity ratio with respect to the number of generated events to determine whether it reaches a plateau, making it unnecessary to perform additional simulations.

## 5. A Mean Projector as a solution to reduce noise

The stochastic nature of Monte-Carlo simulations induces a statistical noise. We thought that by reconstructing images with a mean projector, probability values may be smoothed and we could reconstruct images with better homogeneity and thus with less noise. In fact, let's assume that in order to calculate the projector we perform four simulations Sim1, Sim2, Sim3 and Sim4 and that we calculate four projectors P1, P2, P3 and P4 using respectively each of these simulations. The difference in the calculation of a projector element $P_{ij}$ performed once by appending the four sub-simulations (P_initial) and another time by calculating a mean of these four sub-simulations (P_mean) is shown in the table 1.

| | P1 | P2 | P3 | P4 | P_initial | P_mean |
|---|---|---|---|---|---|---|
| Number of photons $N_{ij}$ detected in the pixel j and emitted from voxel i = 1, 2, 3 and 4. | $x_1$ | $x_2$ | $x_3$ | $x_4$ | $\sum_{i=1}^{i=4} x_i$ | / / |
| Number of photons emitted from voxel i | $n_1$ | $n_2$ | $n_3$ | $n_4$ | $\sum_{i=1}^{i=4} n_i$ | / / |
| Projector element $P_{ij}$ | $\frac{x_1}{n_1}$ | $\frac{x_2}{n_2}$ | $\frac{x_3}{n_3}$ | $\frac{x_4}{n_4}$ | $\frac{\sum_{i=1}^{i=4} x_i}{\sum_{i=1}^{i=4} n_i}$ | $\frac{1}{4} \times \sum_{i=1}^{i=4} \frac{x_i}{n_i}$ |

**Table 1. Calculation of projector's elements**

Obviously, we can notice that except the case where $n_1 = n_2 = n_3 = n_4$, $P_{ij}$ has different values in P_initial and P_mean. In our case, we have used the results of 460 sub-simulations in order to calculate a set of 20 projectors each one calculated from a set of 23 sub-simulations; we obtain so a set of 20 sample-projectors Pi ($i \in [1 \ldots 20]$).

## 6. Bootstrapping as an alternate issue to reduce noise

Statistically, it is known that the standard deviation is an inverse function of the number of measures. Each of the projectors Pi ($i \in [1 \ldots 20]$) calculated above can be considered as a sample projector. By re-sampling, bootstrapping generates a new set of samples and increases the statistics included in the calculation of the projector estimate and thus reduces the variance of estimated parameters [Efron et al. 1979].

Initial Projector:

| | | | | | | |
|---|---|---|---|---|---|---|
| Simulation1 | Simulation2 | Simulation3 | ….. | Simulation i | ..... | Simulation460 |

Sample Projectors:

| P1 | P2 | …. | Pi | …. | P20 |
|---|---|---|---|---|---|
| Simulation1 | Simulation24 | | Simulation[(i-1)x23]+1 | | Simulation438 |
| Simulation2 | Simulation25 | | … | | … |
| …. | …. | | | | |
| Simulation23 | Simulation46 | | Simulation(ix23) | | Simulation460 |

| | | | | |
|---|---|---|---|---|
| P1 | P2 | P3 | P3 | P3 |
| P7 | P9 | P10 | P11 | P11 |
| P14 | P14 | P15 | P17 | P18 |
| P18 | P18 | P18 | P19 | P20 |

**Bootstrap sample 1 (sorted by increasing i)**

……….. ……

……….. …...

| | | | | |
|---|---|---|---|---|
| P7 | P3 | P5 | P8 | P6 |
| P6 | P8 | P7 | P5 | P9 |
| P13 | P12 | P13 | P14 | P16 |
| P17 | P11 | P18 | P12 | P19 |

**Bootstrap sample 100 (sorted randomly)**

**Figure 4. Re-sampling projectors**

$$P_bootstrapped = \frac{\sum_{i=1}^{i=100} Bootstrap_sample_i}{20 \times 100}$$

On figure 4, we see how a sample projector Pi is computed with a set of 23 sub-simulations, whereas each bootstrap sample is a set of 20 randomly selected sample projectors (Pi). Drawings are done with replacement so we can have many identical Pi in the same bootstrap sample.

The bootstrap projector was then the mean of 100 bootstrap sample projectors. Results of relative quantitation are shown in figure 9. Reference values are plotted in the last column .We plotted relative quantitation values for each rod calculated on images reconstructed with the initial projector and the bootstrapped projector (-B refers to a bootstrapped projector). Curves obtained by the initial projector and bootstrapped projector are almost superposed.

## Results

In order to study the change of relative quantitation with respect to the number of generated events, we calculated projectors corresponding to different statistics going from a projector calculated from 33 sub-simulations to 462 sub-simulations with a step of 33 simulations. We then reconstructed the image using each of the resulting 14 calculated projectors: P33, P66…P462. The accuracy of the reconstructed images was assessed by measuring the activity ratio in the 6 rods (Fig 5). Results were plotted for each rod (rod's diameter expressed in millimeter). We can see the improvement in accuracy with the increase of generated events. The reference to obtain is given in the last column.

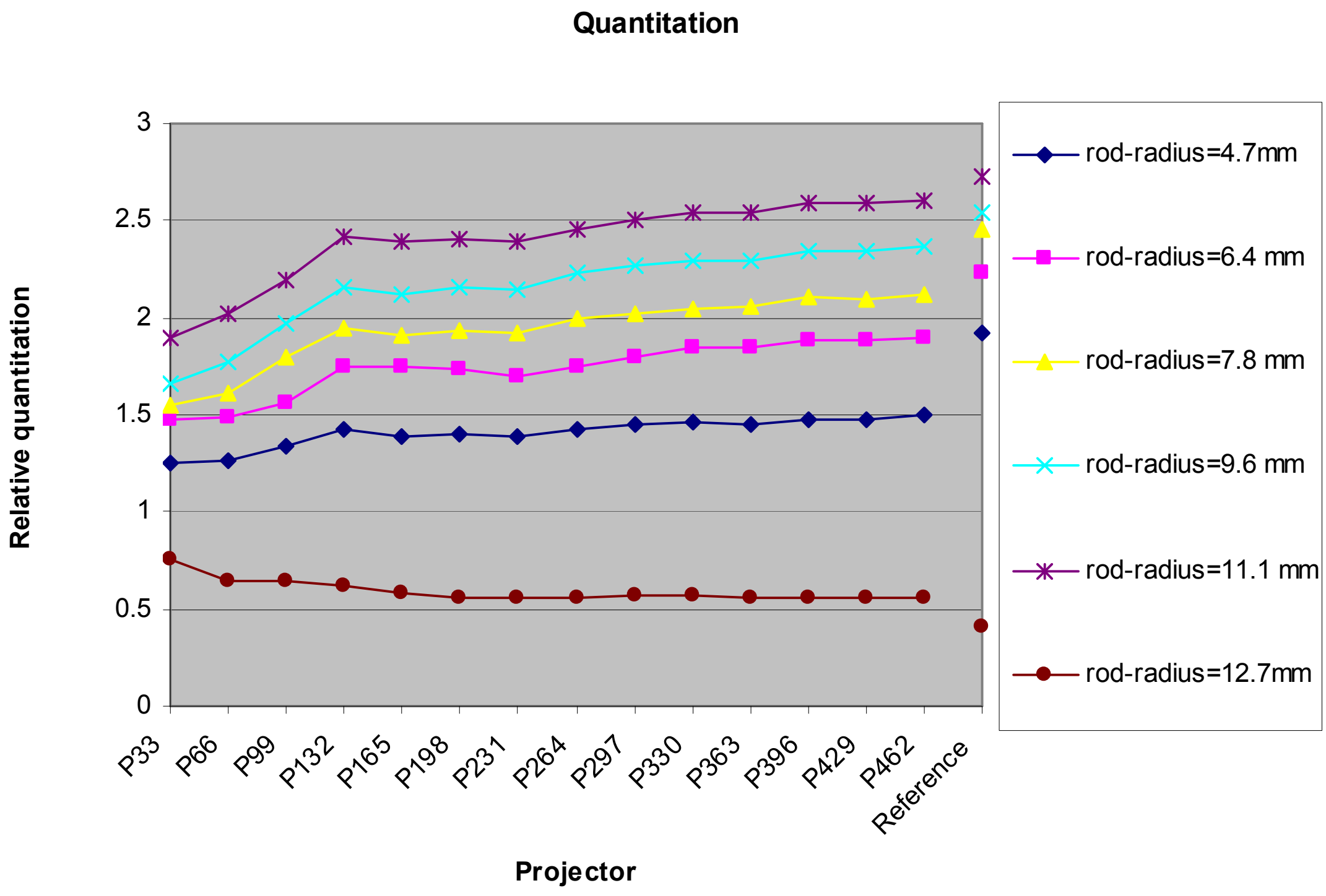

**Figure 5.Relative quantitation with respect to the number of simulation output files**

The mean projector used to reconstruct images was defined as the mean of 20 projectors Pi ($i \in [1..20]$), each of these 20 projectors being calculated from a set of 23 sub-simulations. Figure 6 shows the smoothing impact of the mean-projector calculation on the distribution of computed probabilities mainly in zone of high frequency. Results of quantitative ratios using the mean projector (-M refers a mean projector) are shown in Figure 7. Curves obtained by the initial projector and mean projector are almost superposed.

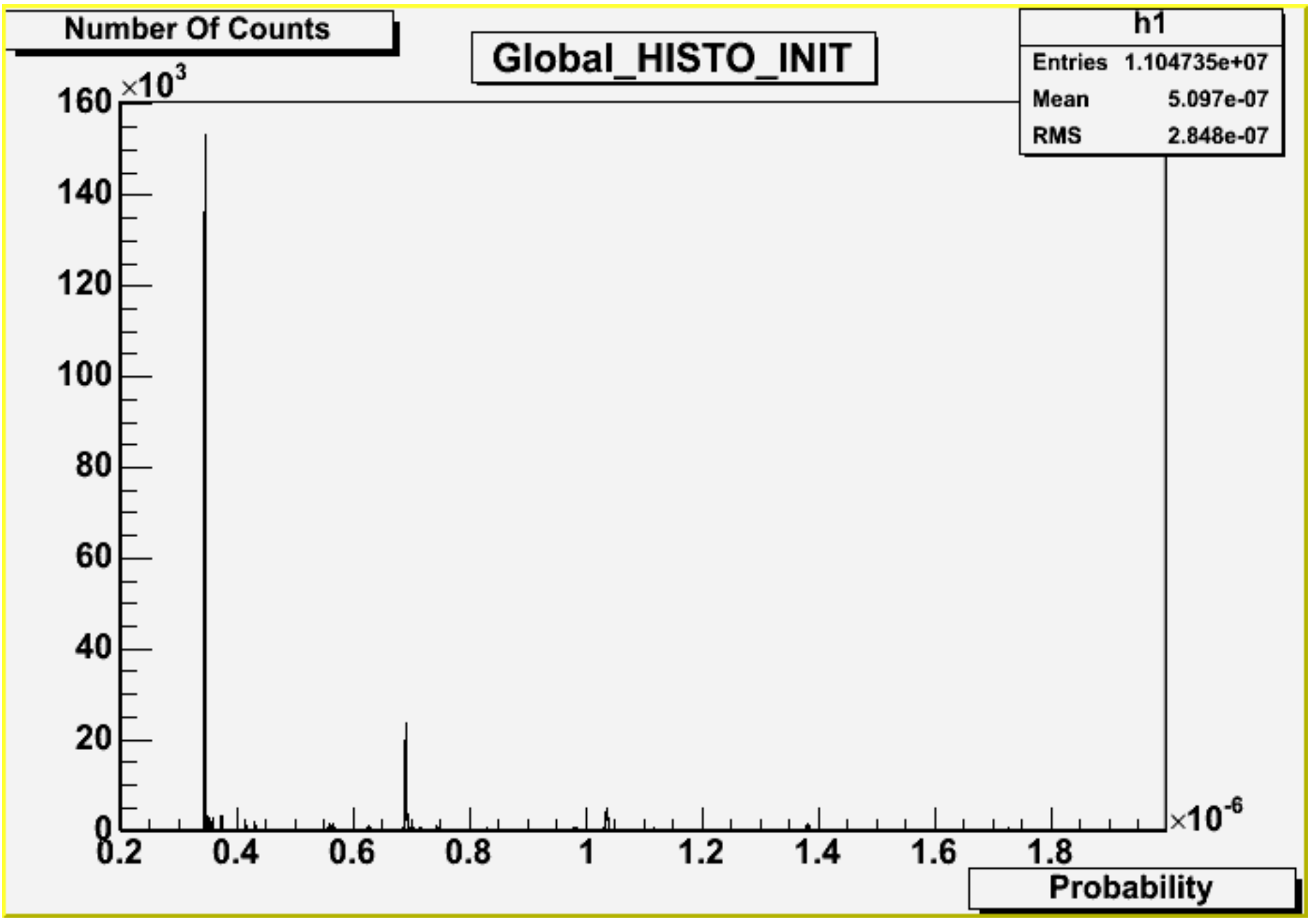

(a)

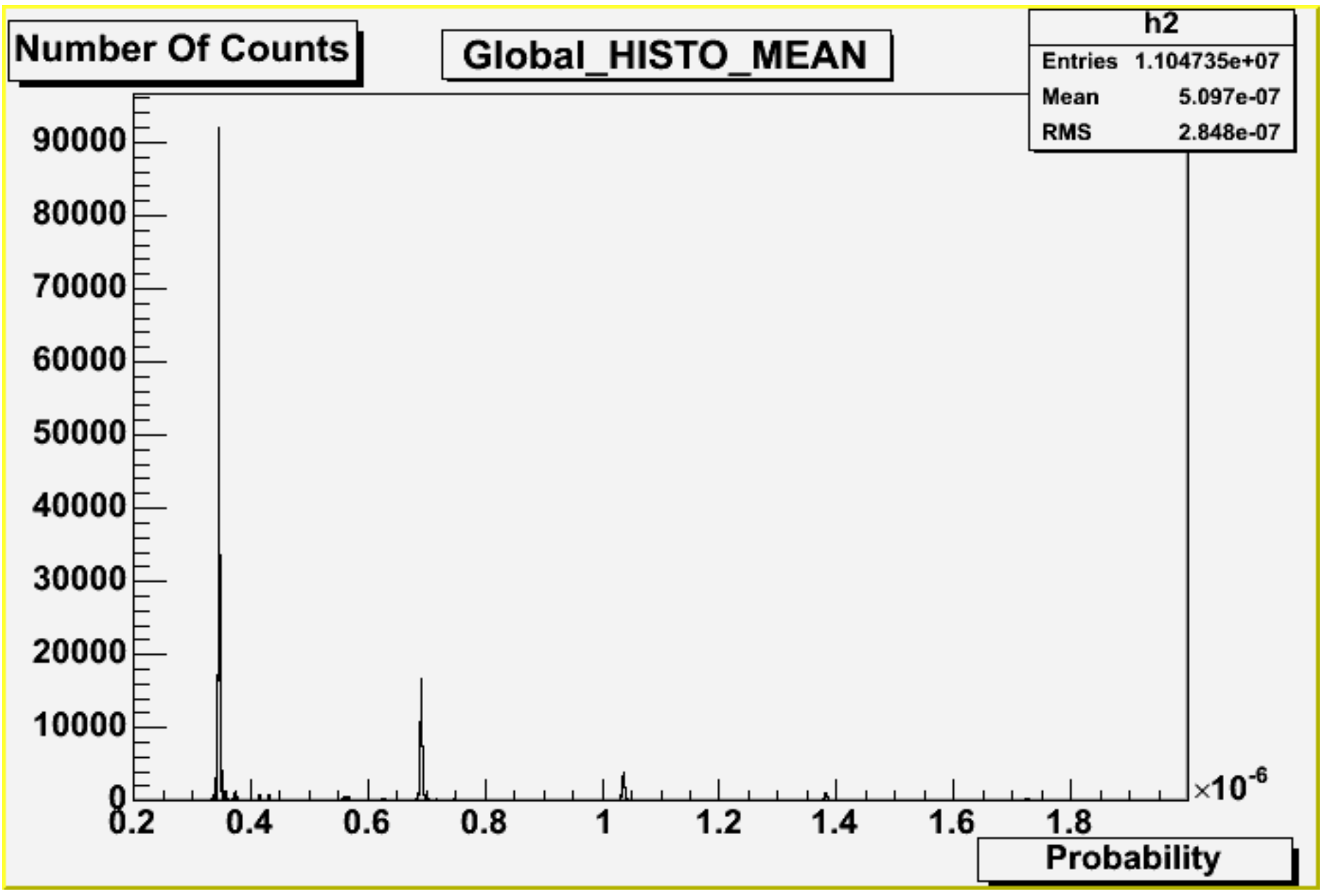

(b)

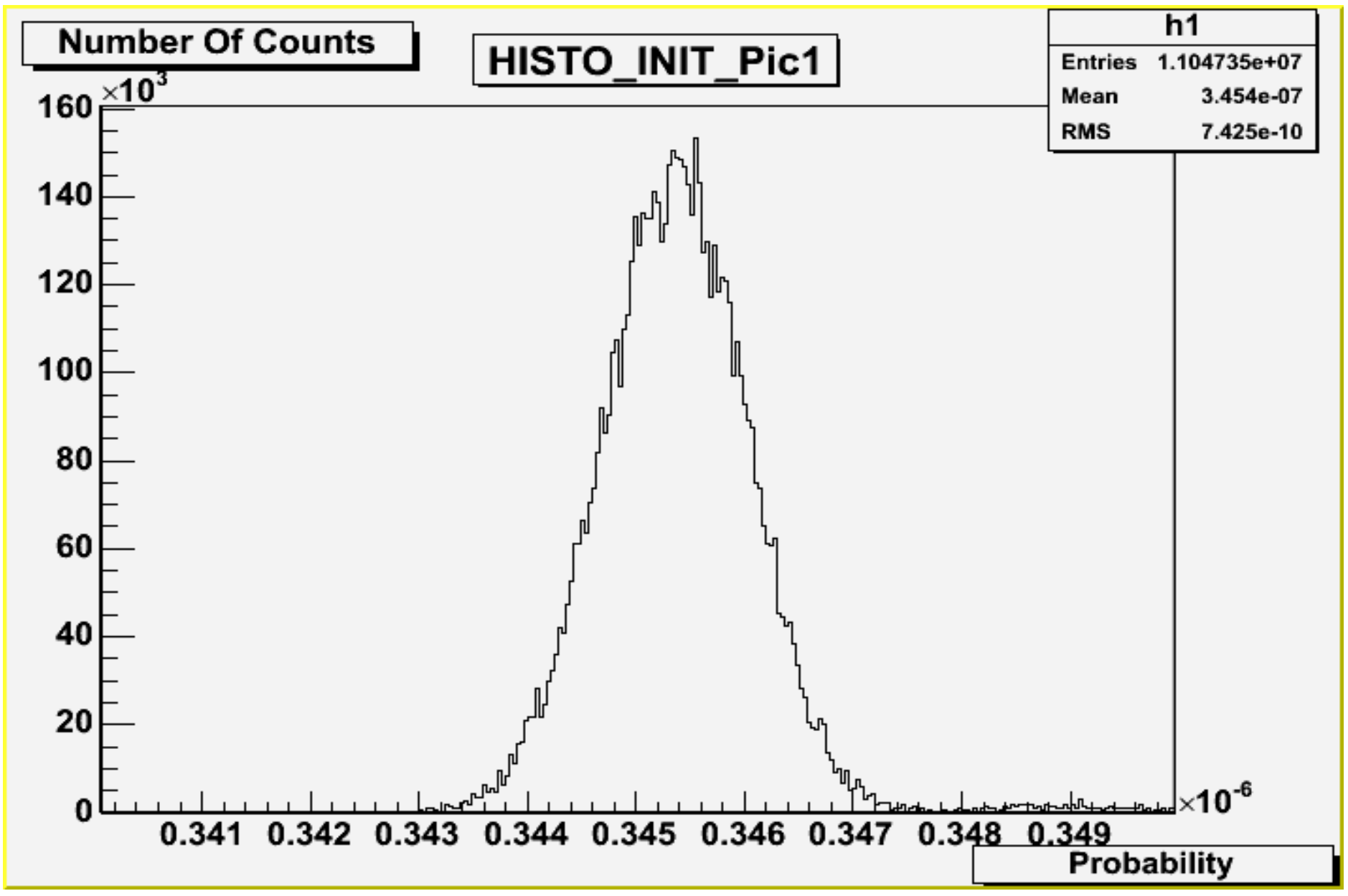

**(c)**

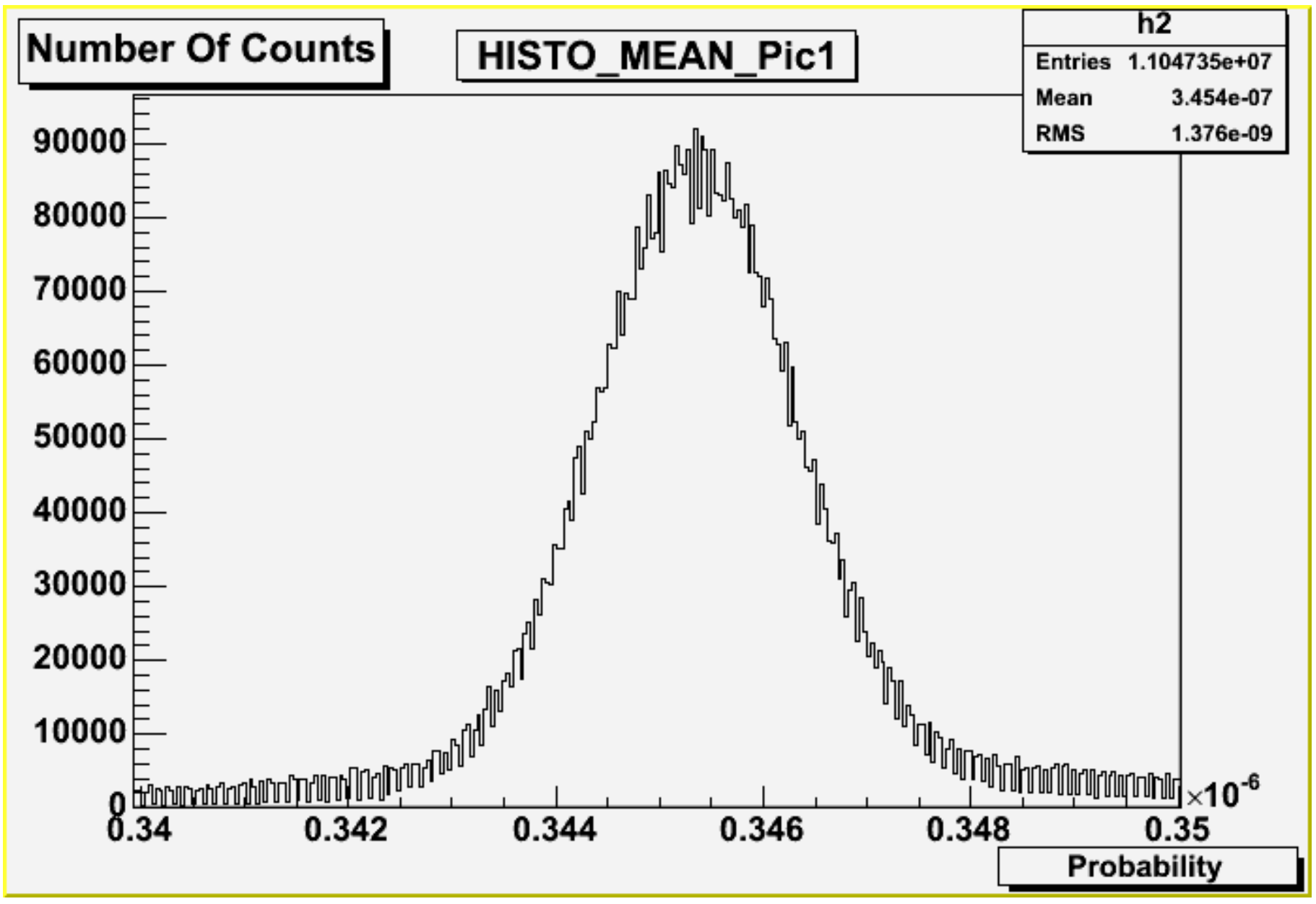

**(d)**

**Figure 6. Probability distribution:**

**(a) Global view of the initial projector ; (b) Global view of the mean projector**

**(c) Zoom on the first peak (initial) ; (d) Zoom on the first peak (mean)**

Difference between initial projector and mean projector are mainly shown in the value of the frequency of the first peak, where we have a value of 160 000 for the initial projector and 90 000 for the mean projector.

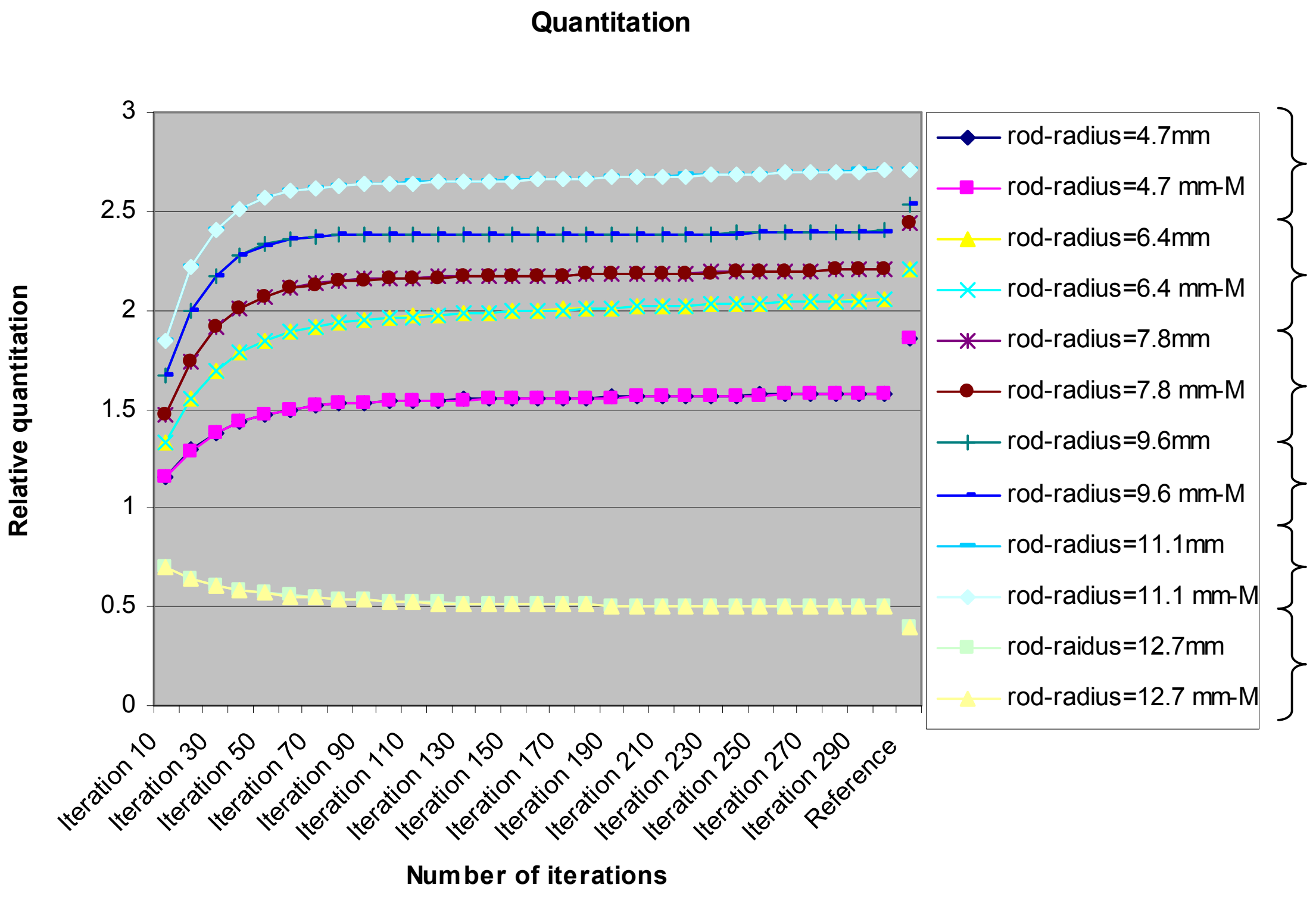

**Figure 7. Comparison between convergence rates of respectively the initial projector and the mean projector**

Instead of calculating the mean projector of 20 elementary projectors Pi (i going from 1 to 20 in increasing order), we drew randomly 20 projectors from the set of the 20 projectors Pi (example: P1, P2, P2, P3, P6, P6….P14, P20) (using drawing with replacement). We repeated this operation 100 times and thus obtained 100 combinations of 20 randomly selected sample projectors, each one forming a bootstrap sample [Chernick et al 1999].(Fig 4)

Figure 8 (a) shows the impact of bootstrapping in the apparition, within the probability spectrum, of new probability values; this could be explained by the fact that bootstrapped projector is obtained by dividing by 2000 the sum of the sub-projectors drawn randomly. The zoom on the zone where a pick appeared in the initial projector and the mean projector (Fig.8.b), shows that bootstrapping induces a quasi-stochastic distribution and that a set of new generated picks was involved within the bootstrapped projector.

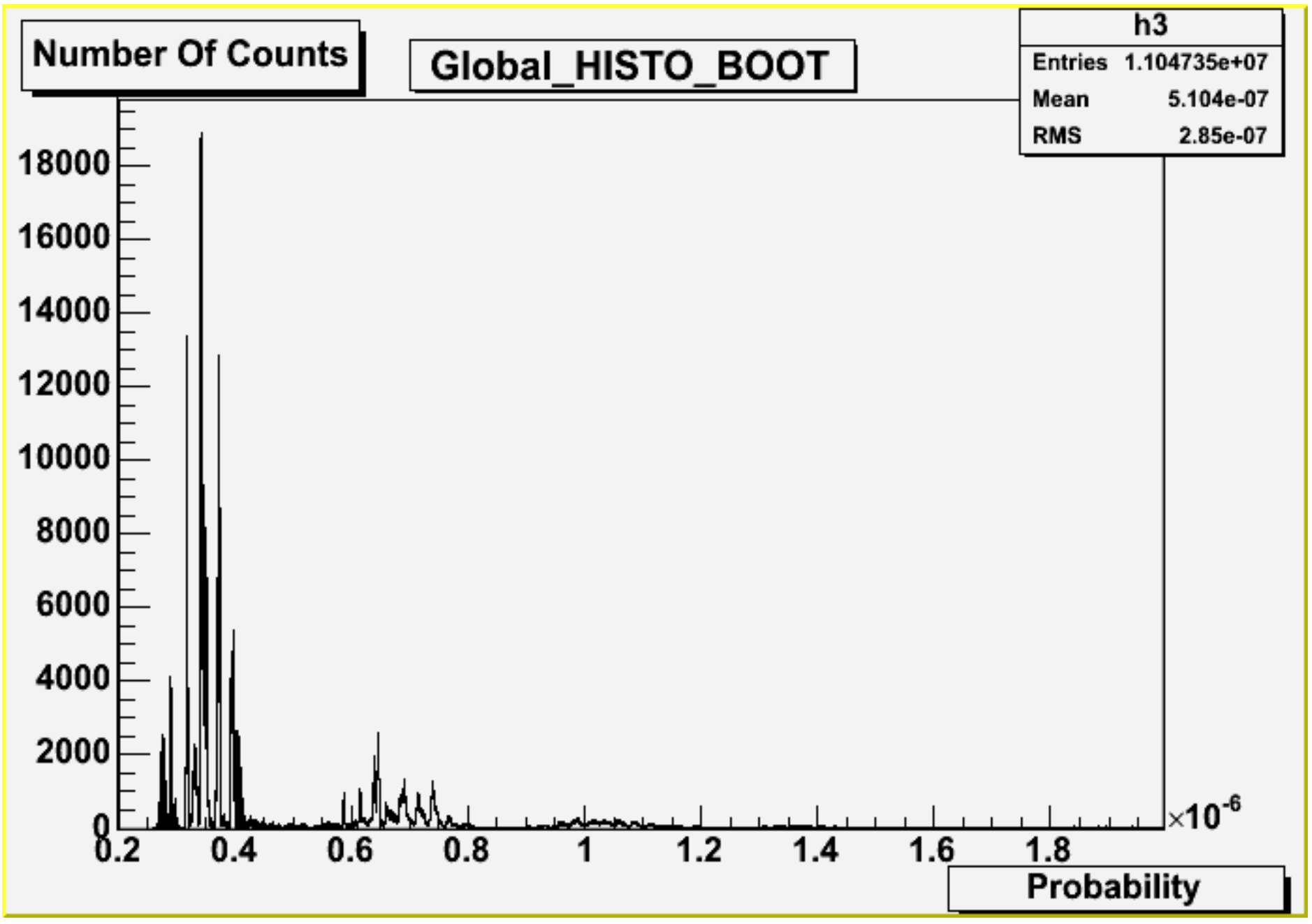

**(a)**

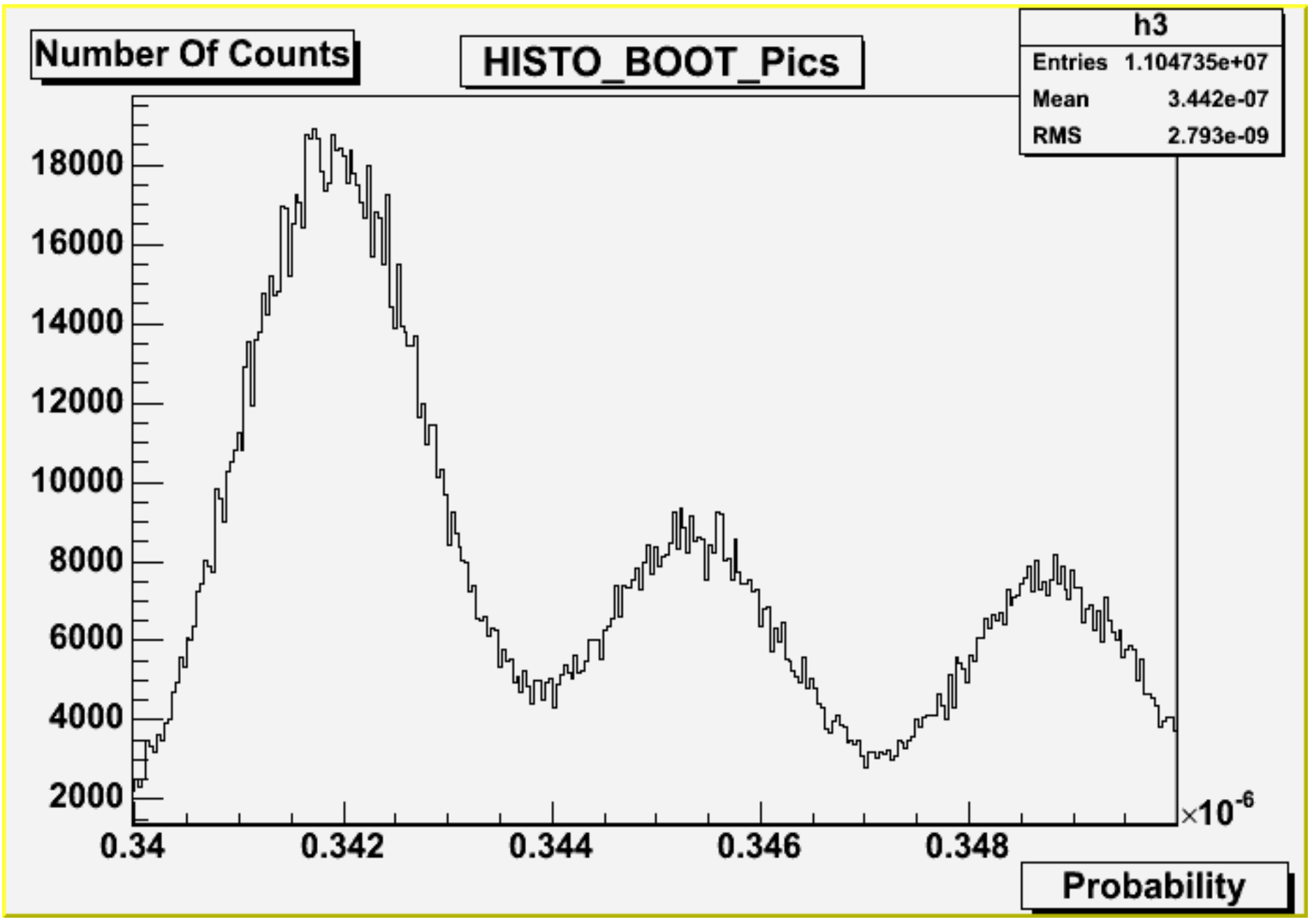

**(b)**

**Figure 8. (a) Global spectrum of the boostrapped projector**
**(b) Zoom on the zone where a pick appears in the initial projector and the mean projector**

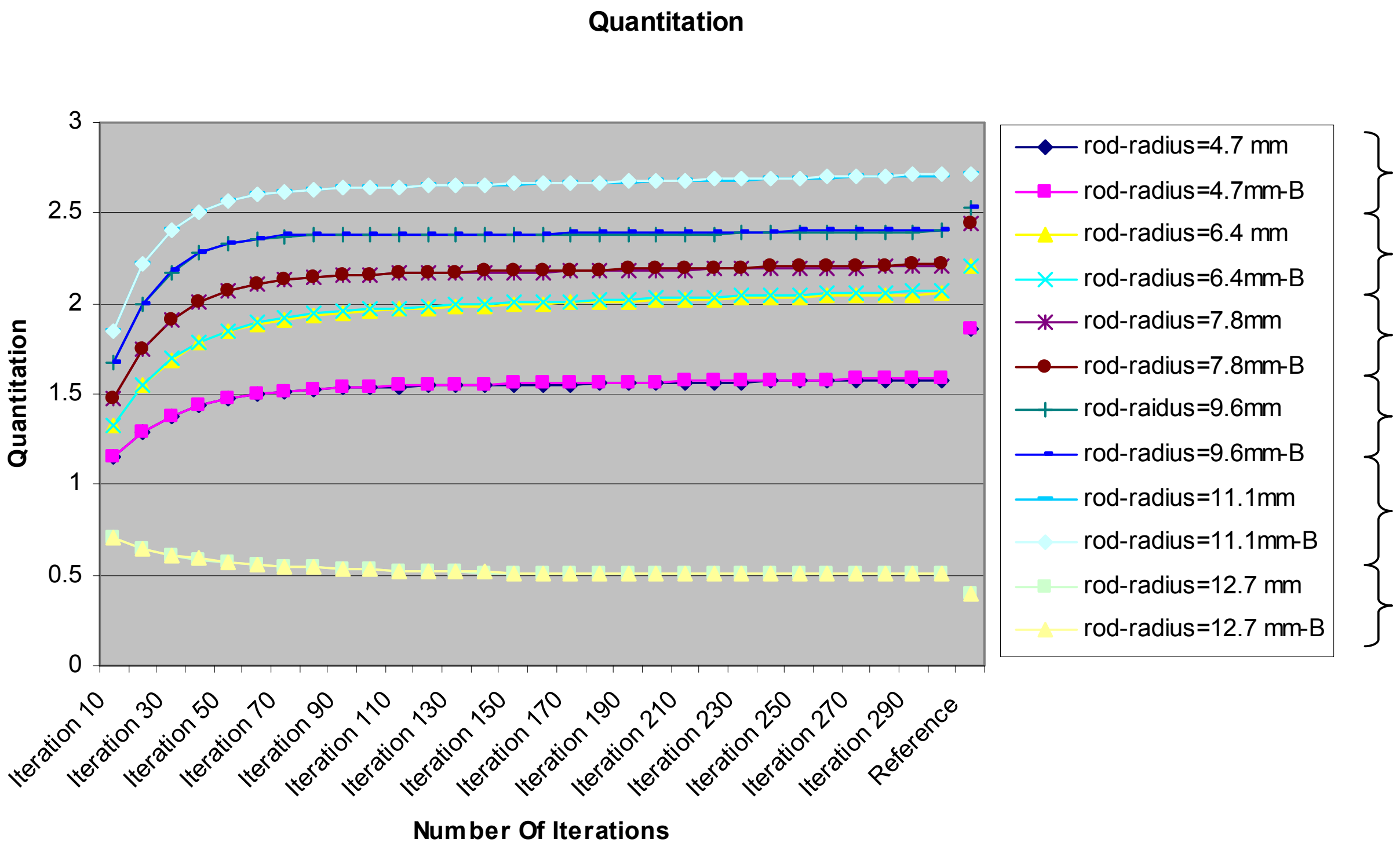

**Figure 9. Comparison between convergence rates of respectively the bootstrapped projector and the initial projector**

The Table 2 shown below presents critical value of probability in different projectors:

| Projector | Minimal Probability | Mean Probability | Maximal Probability | Standard Deviation | Variance |
|---|---|---|---|---|---|
| Initial | $3.421\times10^{-7}$ | $5.41\times10^{-7}$ | $3.01\times10^{-5}$ | $4.762\times10^{-7}$ | $2.267\times10^{-13}$ |
| Mean | $3.257\times10^{-7}$ | $5.41\times10^{-7}$ | $2.983\times10^{-5}$ | $4.764\times10^{-7}$ | $2.27\times10^{-13}$ |
| Bootstrapped | $2.606\times10^{-7}$ | $5.41\times10^{-7}$ | $3.254\times10^{-5}$ | $4.745\times10^{-7}$ | $2.251\times10^{-13}$ |

**Table 2.Critical values in different projectors**

We can easily notice that bootstrap technique has enlarged the interval of probability including new values lower than the minimum probability value and higher than the maximum probability value in the initial projector (i.e. in the mean projector).

## Discussion

Results shown in figure 4 demonstrate that overall, best accuracy is obtained for the largest number of histories considered for calculating the MC projector, as expected from the stochastic nature of MC simulations. We also notice that even if P132 (calculated from 132 sub-simulations) included fewer histories than P231 (calculated from 231 sub-simulations), it tended to give better results than P231, which also reflects the stochastic nature of the calculations involved. Bias between reference values and the plateau could be explained by the fact that the number of simulated events involved in the computation of the projector is still not high enough.

Results shown in figure 4 suggest that initial projector and bootstrapped projector are nearly identical which means that combinations of initial projectors Pi in almost all cases yielded a bootstrapped projector quasi-equal to the initial projector.

The weak decrease of variance in the bootstrapped projector (Table 2) indicates that we may still not have an important statistics involved in the calculation of the projectors (P1… P20) and that we should perform more simulations. This later comment could be a first explanation of the new generated picks in the bootstrapped projector.

A second explanation could be that the projector calculated from the 460 simulations is already a kind of bootstrapped projector since the Independent Sequence technique used seeds selected in an increasing order (not randomly). Indeed the seed initialization was taken in the [1,460] interval. It is known that the IS can lead to encapsulated sub-sequences, since a random number generated in one sub-sequence can match the seed used for another sub-sequence. Some authors warn scientists who run MC experiments in parallel [Hellekalek 1998]. The problem lies in the parallelization of pseudo random numbers. As for many parallelizing techniques of Pseudo Random numbers, the main problem of those methods is the "Long-Range" correlation [De Matteis et al, 1988]. Variants of the known techniques have been developed, a notable one is the parameterization method which is a variant of the IS technique [Srinivasan et al, 1999]. It consists in parameterizing both the seed and iteration function (i.e. the function that gives the next state in the sequence). One of the main contribution of this variant is that it results in a scalable period.

A way to confirm this hypothesis is to have a simulation where we simulate a number of histories equal to the number of histories simulated within 460 separated simulations. Once this simulation is performed, we can compare the results of the quantitative ratio with those obtained by performing 460 separated simulations with independent seeds. However, we are still investigating in order to understand the spectrum of bootstrapped projector.

## Conclusion and Perspectives:

This paper is placed in the context of tomographic reconstruction in Single Photon Emission Computed Tomography (SPECT). The projector used for reconstruction is obtained by Monte Carlo Simulation. Thus we have studied the impact of the number of generated events on the accuracy of the projector. We studied various issues in order to improve its accuracy either by computing a mean projector or by applying the bootstrap technique which has been detailed above.

Even if results were not satisfying so far, investigations are in progress to improve the accuracy of the projector. As a solution to the slowness of GATE simulations, we intend to use faster MC software named SimSet [Lewellen et al. 1998]. With this simulator, we can quickly generate a high number N of events but it relies on some analytical approximations of stochastic laws that could also be source of inaccuracy. This high number of events could then be split into 20 sub-simulations, from which bootstrapping can be implemented as described above. The fact that the variance in the bootstrapped projector hasn't decreased enough leads us to investigate two possibilities: either the number of histories is still not enough or the possibility of an eventual correlation between sub-simulations. Getting rid of possible dependencies between simulations and increasing the number of histories by using a faster simulator (SimSet), bootstrapping might still improve the quality of the projector.

**References :**

[Efron B, 1979], Computers and the Theory of Statistics : *Thinking the Unthinkable in SIAM Review, vol*. 21*, n°* 4*, pp.* 460-480*, octobre* 1979.

[Cheng R.C.H, 2001], Analysis of simulation experiments by bootstrap resampling. *In proceedings of the 2001 Winter Simulation Conference*. B.A Peters, J.S.Smith, D.J.Medeiros, and M.W.Rohrer, eds

[Cheng R.C.H, 1995], Bootstrap methods for computer simulation experiments. *Proceedings of the 1995 Winter Simulation Conference,* edited by C.Alexopoulos, K.Kang, W.R.Lilegdon, and D.Goldsman.

[Chernick, M.R, 1999], *Bootstrap Methods A Pracititioner's Guide.*New York, Wiley.

[Coddington, 1996], Coddington. *Random number generators for parallel computers*. NHSE Review, 1996 Volume, Second Issue.

[De Matteis et al, 1988], De Matteis A., and Pagnutti, S. *Parallelization of random number generators and Long-Range correlations*. Numer. Math., n°53, 1988, pp. 595-608.

[Davison,A.C and D.V. Hinkley, 1997], *Bootstrap Methods and Their Application.* Cambridge University Press.

[Friedman, L.W. and H.H. Friedman, 1995], Analyzing simulation output using the bootstrap method. *Simulation,* 64, n◦.2, February 1995, pp.95-100.

[Jan et al, 2004], GATE:a simulation toolkit for PET and SPECT *Phys.Med.Biol.*49:4543-61

[Hellekalek 1998], Hellekalek, P. *Don't trust parallel Monte Carlo !.* Proceedings of the 12th workshop on Parallel and distributed simulation. Banff Canada. May 26 - 29, 1998, pp. 82-89.

[Lazaro D et al, 2004], Breton V. and Buvat I. Feasibility and value of fully 3D Monte Carlo reconstruction in Single Photon Emission Computed Tomography. *Nucl. Instr. and Meth. Phys. Res.* **A 527** : 195-200

[Lewellen TK et al, 1998], Harrison RL and Vannoy. The SimSET program. In: MonteCarlo calculations in nuclear medicine. Applications in diagnostic imaging. M. Ljungberg, S. E. Strand and M.A. *King editors. Bristol, Institute of physics publishing,* 1998

[Miller Mi et al, 1985], Snyder DL and Miller TR . Maximum-Likelihood reconstruction for Single-Photon Emission Computed-Tomography. *IEEE Tr.Nucl.Sci.***NS-32:**769-778.

[Munley M T et al, 1991], Floyd C E, Tourassi G D, Bowsher J E and Coleman R E 1991 Out-of-plane photons in SPECT *IEEE Trans.Nucl.Sci.* **38** 776-9.

[Srinivasan et al, 1999], Srinivasan, A, Ceperley, D. and Mascagni, M. *Testing parallel random number generators*. Proceedings of the 3rd Int. Conf. on Monte Carlo and Quasi Monte Carlo Methods in Scientific Computing. J. Spanier et al. (ed.), Springer-Verlag, Berlin, 1999.